\documentclass[
    reprint,         
    aps,             
    pra,             
    superscriptaddress, 
    showpacs,        
    floatfix,        
    nofootinbib      
]{revtex4-2}

\usepackage{amsmath}
\usepackage{amssymb}
\usepackage{amsthm}
\usepackage{bbold,bm}
\usepackage[cal=boondox]{mathalfa}
\usepackage{graphicx}
\usepackage{mathtools}
\usepackage{soul}
\usepackage{ulem}

\usepackage{hyperref}

\usepackage[T1]{fontenc}
\usepackage[utf8]{inputenc}

\newcommand\vex[1]{\mathbf{#1}}

\def\bra#1{\mathinner{\langle{#1}|}}
\def\ket#1{\mathinner{|{#1}\rangle}}

\def\tr{\mathrm{tr}}

\usepackage{xcolor}

\newcommand{\extra}[1]{\textcolor{cyan}{}}

\begin{document}

\title{
Quantum simulation of the Haldane phase using open shell molecules
} 

\author{Suman Aich}
\affiliation{Department of Physics, Indiana University, Bloomington, Indiana 47405, USA}

\author{Ceren B. Dag}
\affiliation{Department of Physics, Indiana University, Bloomington, Indiana 47405, USA}
\affiliation{Quantum Science and Engineering Center, Indiana University, Bloomington, Indiana 47405, USA}

\author{H.A. Fertig}
\affiliation{Department of Physics, Indiana University, Bloomington, Indiana 47405, USA}
\affiliation{Quantum Science and Engineering Center, Indiana University, Bloomington, Indiana 47405, USA}

\author{Debayan Mitra}
\affiliation{Department of Physics, Indiana University, Bloomington, Indiana 47405, USA}
\affiliation{Quantum Science and Engineering Center, Indiana University, Bloomington, Indiana 47405, USA}

\author{Babak Seradjeh}
\affiliation{Department of Physics, Indiana University, Bloomington, Indiana 47405, USA}
\affiliation{Quantum Science and Engineering Center, Indiana University, Bloomington, Indiana 47405, USA}

\begin{abstract}

Dipolar molecules in optical traps are a versatile platform for studying 
many-body phases of quantum matter in the presence of strong and long-range interactions. The dipolar interactions in such setups can be enabled by microwave driving  opposite parity rotational levels of the molecules. We find that the regime where the $N=0,J=1/2,F=1$ state is coupled to the $N=1,J=3/2,F=2$ manifold with circularly polarized microwaves, in the presence of a small magnetic field, can lead to spin-1
quantum magnetic Hamiltonians, due to the decoupling between electron spin and orbit, that is unique to the $^2\Sigma$ ground state molecules.  We demonstrate that in one dimension, the phase diagram associated with this
Hamiltonian, computed via tensor network methods, hosts the celebrated Haldane phase. We find that the Haldane phase persists even in the presence of SU(3) correction terms that break the SU(2) algebra of the Hamiltonian. We discuss the feasibility of the proposed scheme for $^2\Sigma$ molecules with large rotational constants such as the directly laser cooled molecule MgF for future experiments. 
\end{abstract}


{
\let\clearpage\relax
\maketitle
}

\section{Introduction}
\label{sec:intro}
Polar molecules trapped in optical potentials have emerged as a powerful and versatile platform for quantum information processing. The existence of a permanent electric dipole moment in the electronic ground state leads to long-range interactions akin to magnetic atoms like Dy \cite{Casotti_Dy_2024} and Er \cite{Su_Erbium_2023} and Rydberg atom arrays \cite{Ebadi_arrays_2021,Manetsch_arrays_2025}. However, it is the complex internal structure of the molecules which enables new opportunities for quantum computation and simulation. For example, the microwave induced dipolar interactions can be tuned to engineer an attractive long-range potential for Cooper-pairing \cite{Cooper_2009_topological_superfluid} or a repulsive potential needed for collisional shielding \cite{Karman_shielding_2018}. These interactions can also be engineered to robustly store quantum information with coherence times exceeding 6.9 seconds in the nuclear spin of a molecule like RbCs \cite{Gregory_RbCs_2021}. Indeed, two-qubit entanglement has been demonstrated between rotational states of two CaF molecules in tweezers with coherence times of around 30~ms \cite {Bao_CaF_entanglement_2023,Holland_CaF_entanglement_2023}. Additionally, low entropy and strongly correlated states of molecules have also been achieved, such as degenerate Fermi gases of KRb \cite{DeMarco_KRb_DFG_2019} and NaK \cite{Schindewolf_NaK_DFG_2022} and a Bose-Einstein condensate of NaCs molecules \cite{Bigagli_NaCS_BEC_2024}. The latter advances further improve the loading of optical lattices and tweezer arrays to achieve higher filling fractions.  

While progress in quantum computation based on atomic and molecular systems is limited by qubit decoherence, duty cycle and system scalability constraints, analogue quantum simulation has emerged as a powerful platform to gain insights into strongly correlated quantum many-body systems and the resulting many-body entangled states. Analysis of such systems can quickly become intractable with classical computational tools with increasing system size~\cite{QSim_review_2021}. This technique of engineering synthetic materials has already been highly successful in elucidating the phase diagram of the Bose-Hubbard \cite{Bakr_Superfluid_Mott_2010,Su_Erbium_2023,Weckesser_Bose_Hubbard_2025,2025arXiv250517009S}, Fermi-Hubbard \cite{Sompet_Fermi_Hubbard_2022,Xu_Fermi_Hubbard_2025} and spin Hamiltonians \cite{Cohen_2015,Semeghini_Rydberg_2021} with both nearest-neighbor (super-exchange) and long-range interactions. Quantum spin Hamiltonians are particularly of interest due to the predicted spin-liquid \cite{Micheli_2006_spin_toolbox} and topological states \cite{Haldane_1983,Cooper_2009_topological_superfluid,2025arXiv250517009S} achievable in certain regimes. However, most studies with dipolar molecular platforms have focused on spin-$\frac{1}{2}$ Hamiltonians, including the \textit{t-J} model with KRb \cite{Carroll_tJ_2025} and the Heisenberg XXZ model with NaRb molecules \cite{Christakis_NaRb_2023}. 

Although spin-$\frac{1}{2}$ particles, such as electrons, describe numerous magnetic phases of matter, additional degrees of freedom, such as  orbital or valley, can lead to higher angular momentum constituents in many-body quantum systems. This is particularly relevant to materials that harbor phases such as Kitaev spin liquids \cite{Chen_Kitaev_spin_liquid_2022}, spiral spin liquids \cite{Gao_spiral_spin_liquid_2017}, Kagome magnets \cite{Haraguchi_Kagome_magnet_2025}, Haldane phases \cite{Tin_Haldane_SPT_2023} and spin-1 Heisenberg antiferromagnetic materials \cite{Buyers_86, Steiner_87, Morra_88, Regnault_94, Sulewski_95, Ma_95, Xu_96, Zaliznyak_01, Kenzelmann_02}.  In this work, we propose a novel quantum simulation scheme that can realize quantum spin Hamiltonians with higher spin, such as spin-1, described by a Hamiltonian satisfying the SU(2) algebra with small SU(3) corrections.
This scheme takes advantage of the complex hyperfine sub-structure of the rotational states in open shell molecules unique to the ubiquitous $^2\Sigma$ radicals. 

\begin{figure}[ht!]
    \centering
    \includegraphics[width=\linewidth]{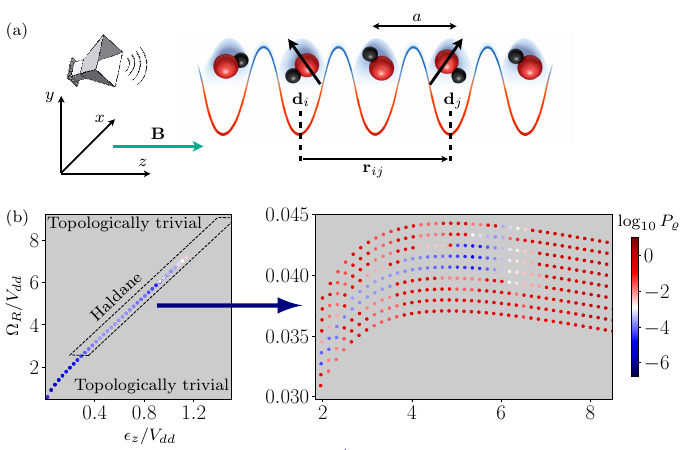}
    \caption{(a) Schematic of the experimental setup showing diatomic dipolar molecules loaded into a one-dimensional lattice of spacing $a$. The depth of the lattice is high such that the molecules are pinned to their sites and hopping is suppressed. The molecules are subjected to circularly-polarized microwaves $\vex E$ of frequency $\omega$ produced by a microwave horn propagating in the $z$-direction, and a static magnetic field $\vex B$, also in the $z$-direction. (b) Phase diagram showing $\log_{10}P_\varrho$ as a function of Zeeman shift $\epsilon_z/V_{dd}$ and Rabi frequency $\Omega_R/V_{dd}$ (bottom left). The bottom right panel shows an expanded view of the rectangular region in the bottom left panel with rotated axes $(\epsilon_z \cos\eta + \Omega_R \sin\eta)/{V_{dd}}$ (horizontal) and $(-\epsilon_z \sin\eta + \Omega_R \cos\eta)/{V_{dd}}$ (vertical), where $\eta = \arccos(1/\sqrt{1+5.45^2})$. The blue data points correspond to the topologically non-trivial Haldane phase and the gray background, and the red data points in the bottom right panel, correspond to topologically trivial phases. In the bottom left panel, red data points have been omitted for the sake of clarity. Using the bottom left panel, the Haldane phase exists for a maximum $\epsilon_z \approx 1.1 \, V_{dd} \approx 55\,\mathrm{kHz}$ ($B \approx 40\,\mathrm{mG}$) and $\Omega_R \approx 7\,V_{dd} \approx 350 \,\mathrm{kHz}$}
    \label{fig:ExpPhaseDiag}
\end{figure}

In particular, our proposed cold molecular platform is composed of a one-dimensional array of open-shell molecules, Fig. \ref{fig:ExpPhaseDiag}(a), that interacts with a near-resonant microwave drive detuned by $\delta$ and of Rabi frequency $\Omega_R$. Simultaneously, a static magnetic field Zeeman shifts the transition by $\epsilon_z$. The resulting microwave dressed molecules interact via dipole-dipole interactions with strength  $V_{dd} \approx 50$~kHz for nearest neighbor molecules. We write down the interaction Hamiltonian in terms of Gell-Mann matrices, which reveals a spin-1 XXZ Hamiltonian together with purely SU(3) interaction terms that present deviations from the SU(2) algebra of spin-1 XXZ model, see Eq.~\eqref{eq:H_eff} below. 

Using the density matrix renormalization group (DMRG) algorithm for finite matrix product states (MPS), we obtain the phase diagram for the XXZ Hamiltonian and show that the symmetry protected topological (SPT) Haldane phase emerges, flanked by topologically trivial XY, antiferromagnetic (AFM) and ferromagnetic (FM) phases. Importantly, we find that even with the inclusion of SU(3) interaction terms in the Hamiltonian, the Haldane phase is stabilized due to the preserved bond-centered inversion symmetry \cite{Pollmann_2010}. This leads to a completely doubly-degenerate entanglement spectrum in the many-body ground state. 

We show the region in the $\epsilon_z$-$\Omega_R$ phase diagram occupied by the Haldane phase in Fig.~\ref{fig:ExpPhaseDiag}(b). To demonstrate the double-degeneracy of the entanglement spectrum, we employ the quantity $P_\varrho = \sqrt{\sum_{q\geq 1} |\varrho_{2q-1} - \varrho_{2q}|^2}$, where $\varrho_q$ are the Schmidt decomposition coefficients of the many-body ground state. In the Haldane phase, $P_\varrho\rightarrow 0$ \cite{Korbmacher_2025}. Using a finite MPS with chain size $L=200$ and a red detuned microwave with $|\delta| = 2\,V_{dd}\approx 0.1$~MHz, we find that the Haldane phase is indeed realizable with bosonic $^{24}$Mg$^{19}$F molecules for accessible experimental parameters
of up to $\epsilon_z \approx 1.1 \, V_{dd} \approx 55$~kHz (static field $B \approx 40$~mG) and $\Omega_R \approx 7\,V_{dd} \approx 350$~kHz.
Thus, our theory presents a novel platform for simulating SPT ordered phases and quantum magnetism of higher spins.     

This paper is organized as follows: In Sec.~\ref{sec:coupling}, we explain the molecular energy level diagram and the experimental conditions that give rise to spin-1 quantum magnetic Hamiltonians, which are derived in the following Sec.~\ref{sec:hamiltonian}. Sec.~\ref{sec:phasediagram} uses the string order parameters and entanglement spectrum to map the phase diagram. Then in Sec.~\ref{sec:properties}, we discuss the experimentally accessible regions of the phase diagram. We conclude in Sec.~\ref{sec:conclusion} with comments on future directions.

\begin{figure}
    \centering
    \includegraphics[width=\linewidth]{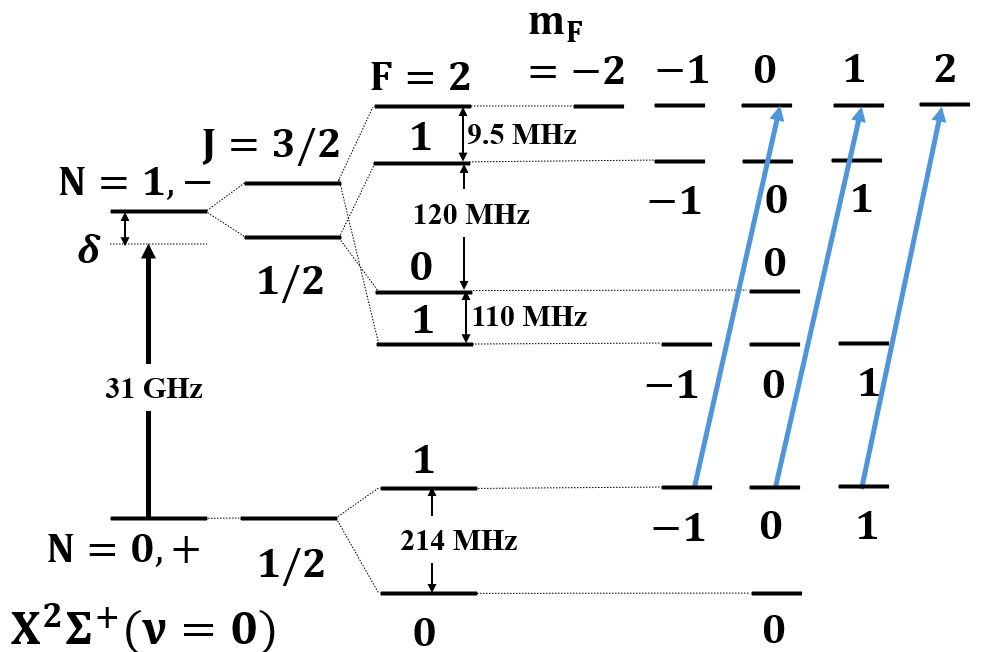}
    \caption{Relevant rotational and hyperfine structure of a $^2\Sigma$ molecule. The lowest rotational state is $N=0$ and has a parity of +, while the first excited rotational state is $N=1$ of the opposite parity. The energy difference between them is $\omega_0$. The sublevel structure is denoted by the quantum numbers $\ket{J,F,m_F}$ leading to the states depicted. The microwave drive is right circularly polarized ($\sigma^+$), is detuned from the transition by $\delta$ and leads to a Rabi frequency of the drive $\Omega_R$. Although the analysis is applicable to any $^2\Sigma$ molecule, we perform numerical simulations for the $^{24}$MgF molecule and the frequencies are indicated for this molecular species \cite{Doppelbauer_MgF_2022}.}
    \label{fig:LevelStructure}
\end{figure}

\section{Microwave-dressed molecular states}
\label{sec:coupling}
We first consider a single molecule irradiated with circularly polarized microwaves at frequency $\omega$. The microwave frequency is chosen to be nearly resonant with the ground state rotational transition between $\ket{N=0,+}$ and $\ket{N=1,-}$ states, which are separated by an energy difference $\omega_0 \sim 31$~GHz as shown in Fig.~\ref{fig:LevelStructure}. The rotational manifold is split into spin-rotation states, which are further subdivided into hyperfine states. In the ground state of a $^2\Sigma$ molecule, the unpaired electron spin is decoupled from the orbital degree of freedom. This implies that ground state $g$-factor nearly equals that of a free electron, and moreover makes ($F$, $m_F$) good quantum numbers at very low magnetic fields. 

In the absence of an external magnetic field, the $\ket{N=0,F=1,m_F=0,\pm1}$ states are all degenerate, as are the $\ket{N=1,F=2,m_F=0,\pm1,\pm2}$ states. Each molecule has a permanent ground state dipole moment $\vex d_0 = d_0(\sin\theta\cos\phi, \sin\theta\sin\phi, \cos\theta)$, where $\theta$ and $\phi$ are the polar angles in three spatial dimensions. We consider the situation where this is coupled to a circularly polarized microwave field $\vex{E}(t)$ with amplitude $E_0$, frequency $\omega$ and detuning $\delta = \omega-\omega_0$. We assume $|\delta|\ll\omega, \omega_0$. As a result, the microwave drives electric dipole transitions between the $N=0$ and the $N=1$ manifolds. Due to selection rules, only three transitions are allowed: $\ket{0,1,-1}\rightarrow\ket{1,2,0}$, $\ket{0,1,0}\rightarrow\ket{1,2,1}$ and $\ket{0,1,1}\rightarrow\ket{1,2,2}$. For each of the three transitions, the time-dependent light-matter Hamiltonian is a two-level system and is given by,
\begin{equation}\label{eq:MolHam}
    \hat H(t) = \hat H_0 + \vex{\hat d}_0 \cdot \vex{\hat{E}}(t)
\end{equation}
in the lab frame, where $\hat H_0$ is the bare molecular Hamiltonian. 
However, due to a difference in Clebsch-Gordon coefficients, the electric dipole transition matrix elements differ among the three transitions. The Hamiltonian Eq.~\eqref{eq:MolHam} can be diagonalized by performing a unitary rotating wave transformation, to obtain the microwave dressed states and their respective energies. 

In the presence of an external magnetic field, the $\ket{N=0,F=1,m_F=0,\pm1}$ states as well as the $\ket{N=1,F=2,m_F=0,\pm1,\pm2}$ states are Zeeman shifted by $g\epsilon_z$, where $\epsilon_z = \mu_BB$. For the transitions $\ket{0,1,-1}\rightarrow\ket{1,2,0}$, $\ket{0,1,0}\rightarrow\ket{1,2,1}$ and $\ket{0,1,1}\rightarrow\ket{1,2,2}$, the $g$-factors are 2.0023, 1.0007 and 0 respectively.
Thus, under the action of a circularly polarized microwave and a static magnetic field, we obtain the dressed states in the lab frame with forms
\begin{align}
    \ket{+}_n & = a_n\ket{N=0,F=1,m_F=n-2} \nonumber \label{eq:plus} \\
             & \qquad {} + b_n e^{-i\omega t} \ket{N=1,F=2,m_F=n-1}, \\
    \ket{-}_n & = b_ne^{i\omega t}\ket{N=0,F=1,m_F=n-2} \nonumber \label{eq:minus} \\
               &\qquad {} - a_n \ket{N=1,F=2,m_F=n-1},
\end{align}
for $n = 1,2,3$. Here 
\begin{equation}
    a_n = -\frac{A_n}{\sqrt{A_n^2 + (\Omega_R/x_n)^2}}, b_n = \frac{\Omega_R/x_n}{\sqrt{A_n^2 + (\Omega_R/x_n)^2}},
\end{equation}
where $(x_1, x_2, x_3) = (\sqrt{6}, \sqrt{2}, 1)$, $\Omega_R = d_0E_0/\sqrt{3}$
is the Rabi frequency and $A_n = \left(\frac{\delta - (3-n)\epsilon_z}{2}\right) + \sqrt{\left(\frac{\delta - (3-n)\epsilon_z}{2}\right)^2 + \left(\frac{\Omega_R}{x_n}\right)^2}$. The dressed state energies are 
\begin{align}
&\varepsilon_{1}^{(\pm)} = \mp\frac{\omega}{2} \pm \left(\sqrt
{\left(\frac{\delta}{2}-\epsilon_z\right)^2 + \frac{\Omega_R^2}{6}} \mp \epsilon_z\right) , \label{eq:quasienergy1} \\
&\varepsilon_{2}^{(\pm)} = \mp\frac{\omega}{2}  \pm  \left(\sqrt
{\left(\frac{\delta-\epsilon_z}{2}\right)^2 + \frac{\Omega_R^2}{2}} \pm \frac{\epsilon_z}{2}\right), \label{eq:quasienergy2}\\
& \varepsilon_{3}^{(\pm)} = \mp\frac{\omega}{2}  \pm  \left(
\sqrt{\frac{\delta^2}{4} + \Omega_R^2} \pm 2\epsilon_z\right).\label{eq:quasienergy3}
\end{align}
We focus on the case $\Omega_R, \epsilon_z,|\delta|\ll\omega,\omega_0$. 
The details of the calculation can be found in Appendix \ref{app:mol_Ham}.

\begin{figure}
    \centering
    \includegraphics[width=\linewidth]{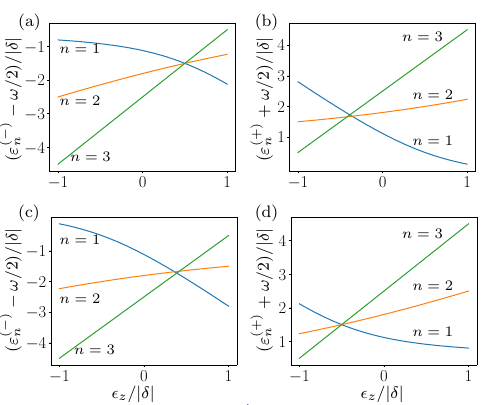}
    \caption{Microwave dressed energy levels $\varepsilon^{(-)}_n - \omega/2$ (left) and $\varepsilon^{(+)}_n + \omega/2$ (right) in Eqs.~\eqref{eq:quasienergy1}-\eqref{eq:quasienergy3}, as functions of the Zeeman shift $\epsilon_z$ for fixed Rabi frequency $\Omega_R = 2.45\,|\delta|$. The top (bottom) panels correspond to a blue (red) detuned microwave $\omega > \omega_0$ ($\omega < \omega_0$).
    }
    \label{fig:Quasienergy}
\end{figure}

Since the microwave dressed states are superpositions of $\ket{N,F,m_F}$ and $\ket{N',F',m'_F}$ with $|N - N'| = |F - F'| = 1$, they have an induced dipole moment with matrix elements $[\vex d]_{nn'}^{\sigma\sigma'} = \prescript{}{n}{\bra{\sigma}}\vex{\hat d}_0\ket{\sigma'}_{n'}$, in the dressed state basis with $\sigma,\sigma' = \pm$. As a result, the microwave dressed molecules can interact via long-range dipole-dipole interactions. 
An important observation about the relevant energy scales in this problem follows from Eqs.~\eqref{eq:quasienergy1}-\eqref{eq:quasienergy3}: the energy splitting between the states $\ket{+}_n$ and $\ket{-}_n$ is $O(\omega)\sim$ GHz, which is much higher than the dipolar interaction strength $d_0^2/a^3\sim$ kHz of the $^2\Sigma$ molecule, with $a$ being the typical lattice spacing of the optical traps we consider. This means that the scattering process $\ket{-}_n\leftrightarrow\ket{+}_k$ due to the dipolar interaction is negligible, and thus the three $\ket{-}_n$ states are energetically well separated from the $\ket{+}_n$ states. 

Additionally, the Zeeman shift allows us to manipulate the energy splitting between the dressed states within each $\ket{\pm}$ subspace. In particular, the dressed energy levels $\varepsilon^{(\pm)}_n$ can be tuned to near degeneracy, as shown in Fig.~\ref{fig:Quasienergy}.
The $\varepsilon^{(-)}_n$ energy levels (left panels in Fig.~\ref{fig:Quasienergy}) cross at positive values of the Zeeman shift $\epsilon_z$ for both blue ($\omega > \omega_0$) and red ($\omega < \omega_0$) detuned microwaves, whereas the $\varepsilon^{(+)}_n$ energy levels (right panels in Fig.~\ref{fig:Quasienergy}) cross at negative values of $\epsilon_z/|\delta|$. Here the sign convention of $\epsilon_z/|\delta|$ is chosen such that a positive $\epsilon_z/|\delta|$ is created by a magnetic field in the positive $z$-direction. Thus, near degeneracy, the dipolar interactions become the dominant energy scale within each subspace of the $\ket{\pm}_n$ dressed states. 

This observation, combined with the fact that the $\ket{\pm}_n$ states  are energetically isolated from each other, allows us to effectively encode a spin-1 degree of freedom in either $\ket{\pm}_n$ subspace of the molecules. Finally, these molecules can interact via strong, long-range dipolar interactions while being optically trapped in a lattice geometry. For the purpose of this work, we consider  molecules prepared in their rotovibrational ground state in a deep lattice, essentially freezing their motional degrees of freedom.  In this situation  tunneling of molecules between neighboring lattice sites may be ignored.

\section{Dipole-dipole interaction Hamiltonian}
\label{sec:hamiltonian}
The dipole moment $\mathbf{\hat d}_0 = (\hat d_x, \hat d_y, \hat d_z)$ of each molecule is a vector operator with matrix elements $[\vex d]_{nn'} = \prescript{}{n}{\bra{\pm}}\vex{\hat d}_0\ket{\pm}_{n'}$ and the dipole-dipole interaction operator is given by
\begin{equation}\label{eq:Vdd_op}
    \hat V_{dd} = \frac{\mathbf{\hat d}_i \cdot \mathbf{\hat d}_j - 3(\mathbf{\hat d}_i \cdot\mathbf{\hat r}_{ij})(\mathbf{\hat d}_j \cdot\mathbf{\hat r}_{ij})}{r^3}.
\end{equation}
This results in a characteristic dipole interaction strength proportional to $d_0^2 (1 - 3\cos^2\theta_{ij})/r^3$, where $\theta_{ij}$ is the angle between the position vector $\vex r_{ij}$ connecting two dipolar molecules, and the quantization axis set by the direction of propagation of the microwave field. In a one dimensional chain, we can choose $\theta_{ij} = 0$ for convenience, see Fig~\ref{fig:ExpPhaseDiag}(a), while in two dimensions the dipolar interaction matrix elements are tunable with $\theta_{ij}$. As a consequence, both attractive (negative) and repulsive (positive) interactions can be engineered, which are predicted to result in a variety of quantum many-body phases such as $p$-wave topological superfluid, charge density wave and Mott insulator \cite{Cooper_2009_topological_superfluid, Lemeshko_2015, Mendes-Santos_2023}.

The dipole moment matrix elements, $[\vex d(t)]_{nn'}$, are in general time-dependent in the lab frame. However, since the microwave drive frequency satisfies $\omega\gg d_0^2/a^3$, the leading order contribution to $\hat V_{dd}$ in the high-frequency approximation comes from the time-average $\hat V_{dd}^{(0)} = \int_0^T \hat V_{dd}(t)\frac{dt}{T}$ ($T = 2\pi/\omega$). The detailed expressions for $[\vex d(t)]_{nn'}$ and $\hat V_{dd}^{(0)}$ are presented in Appendix \ref{app:Dipole}.

Next we derive the effective spin-1 Hamiltonian of the molecular array in the lab frame within the high-frequency approximation. Since the molecules have a three dimensional Hilbert space, the matrix $\mathbb V_{dd}$ can be written as a linear combination of the tensor products of Gell-Mann matrices including the identity matrix, $\bm\lambda = \{\lambda_0, \lambda_1,\cdots,\lambda_8\}$, with $\lambda_0 = \sqrt{2/3}~\mathbb{1}$, 
satisfying the normalization condition $\tr(\lambda_k\lambda_l) = 2\delta_{kl}$. We then choose a particular SU(2) subalgebra  wherein the spin-1 matrices $\vex S = \{S^x, S^y, S^z\}$ can be written as $S^x = (\lambda_1 + \lambda_6)/\sqrt{2}$, $S^y = (\lambda_2 + \lambda_7)/\sqrt{2}$ and $S^z = (\lambda_3 + \sqrt{3}\lambda_8)/2$. This allows us to decompose $\mathbb{V}_{dd}$ as
\begin{equation}\label{eq:V_dd}
    \mathbb{V}_{dd} = \sum_{\alpha\beta}J_{\alpha\beta}S^\alpha\otimes S^\beta + \tilde{\mathbb{V}},
\end{equation}
where the first term contains bilinear spin-1 operators ($\alpha,\beta = 0,x,y,z$ with $S^0 = \mathbb 1$) and the second term $\tilde{\mathbb{V}}$ contains biquadratic spin-1 terms, as well as purely SU(3) terms. Here $J_{\alpha\beta}$ are the spin-1 exchange couplings.   
The matrices $\mathbb{V}_{dd}$, $S^\alpha\otimes S^\beta$ and $\tilde{\mathbb{V}}$ can also be written as linear combinations of tensor products of Gell-Mann matrices
\begin{align}
    & \mathbb V_{dd} = \sum_{kl} v_{kl} \lambda_k\otimes\lambda_l,\\
    & S^\alpha\otimes S^\beta = \sum_{kl} \Xi^{\alpha\beta}_{kl} \lambda_k\otimes\lambda_l, \\
    & \tilde{\mathbb{V}} = \sum_{kl} \Lambda_{kl} \lambda_k\otimes\lambda_l \label{eq:Lambda_kl},
\end{align}
where the expansion coefficients $v_{kl} = \frac{1}{4}\tr[(\lambda_k\otimes\lambda_l)\mathbb V_{dd}]$, $\Xi^{\alpha\beta}_{kl} = \frac{1}{4}\tr[(\lambda_k\otimes\lambda_l) (S^\alpha\otimes S^\beta)]$ and $\Lambda_{kl} = \frac{1}{4}\tr[(\lambda_k\otimes\lambda_l)\tilde{\mathbb{V}}]$. In order to obtain a dominant SU(2) contribution to $\mathbb V_{dd}$, we minimize the sum of the squares of the expansion coefficients $\Lambda_{kl}$ of $\tilde{\mathbb V}$,
\begin{equation}
   \sum_{kl}|\Lambda_{kl}|^2 = \sum_{kl}\big|v_{kl} - \sum_{\alpha\beta}J_{\alpha\beta}\Xi^{\alpha\beta}_{kl}\big|^2,
\end{equation}
with respect to $J_{\alpha\beta}$,
\begin{equation}\label{eq:Minimize}
    \frac{\partial}{\partial J_{\alpha\beta}}\sum_{kl}|\Lambda_{kl}|^2 = 0.
\end{equation}
This results in a linear matrix equation that can be efficiently solved numerically to obtain the spin-1 exchange couplings $J_{\alpha\beta}$. Interestingly, for a one dimensional chain with $\theta_{ij} = 0$, the minimization procedure results in the vanishing of a majority of the interactions, and the only non-zero interactions are $J_{xx} = J_{yy}$, $J_{zz}$, and $J_{0z} = J_{z0}\equiv J_z$. 

Thus the effective Hamiltonian describing molecules arranged in a 1D chain and interacting via long-range, anisotropic dipolar interactions takes the form of a \mbox{spin-1} XXZ model 
perturbed by higher-order SU(2) and purely SU(3) interaction terms,
\begin{align}\label{eq:H_eff}
    H_{\text{eff}} = & \sum_{i<j}\frac{1}{|i-j|^3}\left[\vphantom{\frac{1}{|i-j|^3}} \frac{J_{xx}}{2}(S^+_iS^-_j + S^-_iS^+_j) + J_{zz}S^z_iS^z_j \right. \nonumber \\
    & \left.\vphantom{\frac{1}{|i-j|^3}} + J_{+-}\left((S^+_i)^2(S^-_j)^2 + \mathrm{h.c.}\right)\right]  \nonumber \\
     & + \sum_{i}\left[ h^z_{i} S^z_i + D(S^z_i)^2 \right] + \mathbb V_{\text{SU(3)}}.
\end{align}
Here $S^\pm = S^x \pm i S^y$, $h_{i}^z = D_z + \sum_{j\neq i} J_z/|i-j|^3$ is the 
transverse field strength, where $D_z = (\varepsilon^{(\pm)}_{1} - \varepsilon^{(\pm)}_{3})/2$, and $D =(\varepsilon^{(\pm)}_{1}-2\varepsilon^{(\pm)}_{2}+\varepsilon^{(\pm)}_{3})/2$ is the single-site anisotropy arising from the energy splitting between the dressed molecular states $\ket{\pm}_{n}$. The SU(3) terms are obtained from the dipolar interaction, and thus contain both one-body and long-range two-body terms, and can be expressed as
\begin{equation}
    \mathbb V_{\mathrm{SU}(3)} = \sum_{i<j}\frac{1}{|i-j|^3}\sum_{k,l\geq1}\Lambda_{kl}\lambda_{k,i}\otimes\lambda_{l,j} + \sum_i\sum_{k\geq1}\Lambda_{k,i}\lambda_{k,i},
\end{equation}
with $\Lambda_{k,i} = \sum_{j\neq i} \Lambda_{k0}/|i-j|^3$. Note that the  transverse field $h^z_i$ and Zeeman shift $\epsilon_z$ are distinct quantities: $h^z_i$ is the effective field generated from the dipolar interaction and the dressed energy splitting, whereas $\epsilon_z$ is due to the external magnetic field that shifts the molecular hyperfine states. 

\begin{figure}
    \centering
    \includegraphics[width=\linewidth]{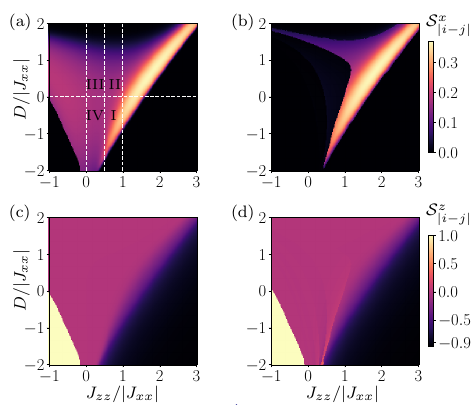} %
    \caption{Phase diagram of $H_{\text{eff}}$ in Eq.~\eqref{eq:H_eff} as a function of $J_{zz}/|J_{xx}|$ and $D/|J_{xx}|$ for $h^z_i = 0$ (left panels) and $h^z_i = 0.1 J_{xx}$ (right panels), with $J_{xx} < 0$ and $\mathbb V_{\text{SU(3)}} = 0$ in both cases. The colormaps show the string order parameters $\mathcal S^x_{|i-j|}$ (bottom panels) and $\mathcal S^z_{|i-j|}$ (top panels) obtained using finite DMRG with a chain size $L = 100$ over a distance of $j - i = 80$, maximum bond dimension $\chi_{\text{max}} = 100$ and the maximum interaction range of 4. Regions I, II, III and IV in (a) denote the experimentally accessible regions in the phase diagram. Table~\ref{tab:ExpParams} summarizes the experimental parameter values for detuning and Zeeman shift, and the initial state of the dipolar molecules required to access these regions.}
    \label{fig:SOP}
\end{figure}

\begin{figure}
    \centering
    \includegraphics[width=\linewidth]{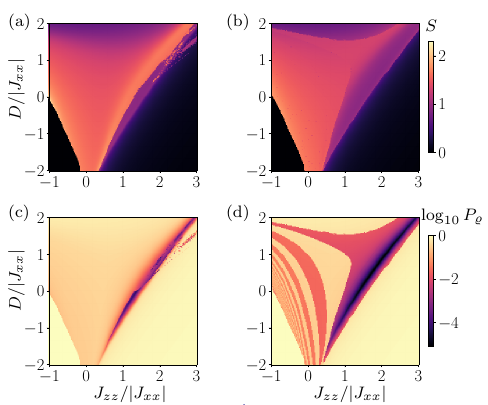} %
    \caption{Phase diagram of $H_{\text{eff}}$ in Eq.~\eqref{eq:H_eff} as a function of $J_{zz}/|J_{xx}|$ and $D/|J_{xx}|$ for $h^z_i = 0$ (left panels) and $h^z_i = 0.1 J_{xx}$ (right panels), with $J_{xx} < 0$ and $\mathbb V_{\text{SU(3)}} = 0$ in both cases. The colormaps show the entanglement entropy $S$ (top panels)  and $P_\varrho$ (bottom panels). The plots are obtained with finite DMRG with the same set of parameters as Fig.~\ref{fig:SOP}.}
    \label{fig:Entanglement}
\end{figure}

We conclude this section by commenting on the origin of $\mathbb V_{\text{SU(3)}}$. Due to the differences in the Clebsch-Gordon coefficients appearing in the transition dipole matrix elements $[\mathbb d_z]_{12}\neq[\mathbb d_z]_{23}$, as well as the diagonal matrix elements $[\mathbb d_x - i\mathbb d_y]_{11}\neq[\mathbb d_x - i\mathbb d_y]_{22}\neq[\mathbb d_x - i\mathbb d_y]_{33}$, the minimization procedure [Eq.~\eqref{eq:Minimize}] results in symmetric and antisymmetric combinations of the matrix elements that contribute to the bilinear spin-1 couplings $J_{\alpha\beta}$ and the SU(3) terms, respectively. Additionally, we also obtain the matrix element $[\mathbb d_x - i\mathbb d_y]_{13}$ that contributes to the biquadratic spin coupling $J_{+-}$. We numerically confirm that $|J_{+-}| \ll |J_{xx}|, |J_{zz}|, |J_z|, |\Lambda_{kl}|$, and so we will drop it from Eq.~\eqref{eq:H_eff} henceforth. 

\section{Phase diagram}
\label{sec:phasediagram}

Now we study the phase diagram of $H_{\text{eff}}$ described by Eq.~\eqref{eq:H_eff} in a one dimensional chain geometry. Initially we fix $\mathbb V_{\text{SU(3)}} = 0$, and consider two cases with and without the transverse field 
$h^z_i$. We obtain the many-body ground state and the resulting phase diagram with finite DMRG using the TeNPy library \cite{Hauschild_2018, Hauschild_2024}.

The nearest-neighbor spin-1 XXZ chain without a transverse field exhibits the Ising ferromagnetic (FM), antiferromagnetic (AFM), Haldane, spin-1 XY and a trivial insulating phase at large $D$ \cite{Haldane_1983, Kennedy_1992, Chen_2003, Mousa_2025}, that persist even under long-range anisotropic interactions \cite{Korbmacher_2025, Brechtelsbauer_2025}. The FM and AFM phases exist for large negative and positive values of $J_{zz}$ respectively, when the Ising-type interactions dominate. Specifically, in the FM phase, all spins occupy $\ket{+1}$ or $\ket{-1}$ states, whereas in the AFM phase, neighboring spins have opposite orientations. Thus, the FM and AFM are gapped phases with long-range order that are characterized by local order parameters such as average magnetization $\langle S^z_i\rangle$ and average staggered magnetization $\langle (-1)^i S^z_i\rangle$, respectively. The long-range order is also characterized by nonzero correlations $\langle S^z_iS^z_j\rangle$ and $\langle (-1)^{|i-j|} S^z_iS^z_j\rangle$ for FM and AFM respectively. For a large positive $D$, the $(S^z)^2$ term dominates and the ground state favors all spins in the $\ket 0$ state, resulting in a trivial insulator. 

For $J_{zz} \sim D$, the ground state may be spin-1 XY or Haldane phases, both of which host large quantum fluctuations. The spin-1 XY is a gapless superfluid phase, and is ferromagnetic (for $J_{xx} < 0$) or antiferromagnetic (for $J_{xx}> 0$). It is characterized by quasi-long-range order (power-law decaying correlations in space) in $\langle S^+_iS^-_j\rangle$ (for $J_{xx}<0$) or $\langle (-1)^{|i-j|}  (S^+_iS^-_j\rangle$ (for $J_{xx}>0$). The Haldane phase in one dimension is a symmetry-protected topological phase with fractionalized spin-$\frac12$ edge states that are four-fold degenerate at zero energy \cite{Gu_2009}. 
It is characterized by nonvanishing string order parameters \cite{PhysRevB.40.4709,PhysRevB.45.304} $\mathcal S^\alpha_{|i-j|} = \langle S^\alpha_i\Pi_{i<k<j}\exp(i\pi S^\alpha_k)S^\alpha_j\rangle$ ($\alpha = x,y,z$) in the limit $|i-j|\rightarrow \infty$, in contrast to the spin-1 XY phase that has a vanishing $\mathcal S^z_{|i-j|}$. 

While the string order parameters $\mathcal S^x_{|i-j|}$ and $\mathcal S^y_{|i-j|}$ are nonvanishing in the XY and Haldane phases, they completely vanish in the FM, AFM and the trivial large-$D$ phases. Additionally, the string order parameter $\mathcal S^z_{|i-j|}$ is nonzero in the Haldane, FM and AFM phases, while vanishing in the XY and the large-$D$ phases. Thus, $\mathcal S^x_{|i-j|}$ or $\mathcal S^y_{|i-j|}$ can be used to distinguish between the Haldane and AFM phases, while $\mathcal S^z_{|i-j|}$ distinguishes between the Haldane and XY phases. The colormaps in the left panels of Fig.~\ref{fig:SOP} show the string order parameters $\mathcal S^x_{|i-j|}$ and $\mathcal S^z_{|i-j|}$ for the various phases with $h^z = 0$.

The topological properties of the Haldane phase are 
protected by certain symmetries, such as 
bond-centered inversion, time-reversal and the dihedral group of $\pi$ rotations about a pair of orthogonal axes, resulting in a doubly-degenerate entanglement spectrum \cite{Pollmann_2010,Pollmann_2012}. The color maps in the left panels of Fig.~\ref{fig:Entanglement} show the entanglement entropy $S = -\sum_q \varrho^2_q \ln\varrho^2_q$ 
and $P_\varrho = \sqrt{\sum_{q\geq 1} |\varrho_{2q-1} - \varrho_{2q}|^2}$, for the various phases with $h^z = 0$. Due to the doubly degenerate entanglement spectrum, $P_\varrho$ vanishes in the Haldane phase.

With a nonzero transverse field along the $z$-axis, one can immediately see that time-reversal and the rotational symmetries $e^{-i\pi S^x}$ and $e^{-i\pi S^y}$ are broken. Nonetheless, the model still has bond-centered inversion symmetry and the product of time-reversal and rotation by $\pi$ about $x$ or $y$, i.e., TR$\times e^{-i\pi S^x}$ or TR$\times e^{-i\pi S^y}$. Importantly, the Haldane phase can be stabilized by the bond-centered inversion symmetry alone \cite{Pollmann_2010}. In this case, the string order parameters are not useful diagnostics to characterize the 
Haldane phase where the string order and the spin-$\frac12$ edge states may be absent \cite{Pollmann_2010,PhysRevB.80.155131,PhysRevB.76.085124}. 

\begin{figure}
    \centering
    \includegraphics[width=\linewidth]{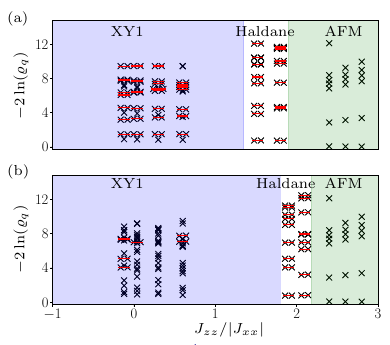}
    \caption{Entanglement spectrum of the many-body ground state of $H_{\text{eff}}$ in Eq.~\eqref{eq:H_eff} as a function of $J_{zz}/|J_{xx}|$ for (a) $h^z_i = 0$ and (b) $h^z_i = 0.5 J_{xx}$, at fixed $D = 0.5 J_{xx}$, with $J_{xx} < 0$ and $\mathbb V_{\text{SU(3)}} = 0$ in both cases. The different phases are shaded with different colors. For degenerate Schmidt values, the crosses are slightly shifted to indicate multiplicity and groups of degenerate Schmidt values are joined by the red line. The plots are obtained with finite DMRG for a chain size of $L = 200$, maximum bond dimension $\chi_{\text{max}} = 100$ and the maximum interaction range of 4.}
    \label{fig:EntSpectrum}
\end{figure}

In this general situation, the double degeneracy of the entire entanglement spectrum arising due to bond-centered inversion, 
is the definitive characteristic feature of the Haldane phase. The colormaps in the right panels of Fig.~\ref{fig:Entanglement} show the entanglement entropy $S$ and $\log_{10}P_\varrho$ for the various phases with $h^z_i = 0.1|J_{xx}|$. We find that in the spin-1 XY phase, $P_\varrho$ oscillates with regions having $P_\varrho \approx 10^{-1} - 10^{-2}$, whereas for $h^z_i = 0$, $P_\varrho$ remains constant. 
In Fig.~\ref{fig:EntSpectrum}(a) and~(b) we show the entanglement spectrum at fixed $D = 0.5|J_{xx}|$ for $h^z_i = 0$ and $h_z=0.5|J_{xx}|$ respectively. For both cases, a narrow region with completely doubly-degenerate spectrum exists, revealing the Haldane phase. On the other hand, the spin-1 XY phase (XY1), has isolated eigenvalues as well as a few doubly degenerate eigenvalues.
In Fig.~\ref{fig:EntSpectrum}, we have increased the system size to $L=200$, to decrease finite-size effects, and find that the oscillatory behavior of $P_\varrho$ in the XY1 persists. Thus we attribute these oscillations  to the presence of 
the transverse field. Finally, we also observe that the Haldane phase shifts towards larger $J_{zz}/|J_{xx}|$ to compensate for the effect of increasing $h^z_i$. 

\begin{table*}[]
\setlength{\tabcolsep}{8pt}
\centering
\resizebox{0.75\textwidth}{!}{%
\begin{tabular}{lll}
\hline\hline
Property & Symbol (unit) & Value \\
\hline
$X^2\Sigma^+$ rotational constant  & $B (2\pi$ MHz)        & 15,496.8125 \\ 
$J=0 \rightarrow J=1$ transition frequency & $\omega_0 (2\pi$ MHz) & 31,010.3    \\ 
$X^2\Sigma^+$ dipole moment        & $d$ (Debye)           & 2.88        \\ 
Nearest neighbor distance     & $a$ (nm)              & 376         \\ 
$A^2\Pi_{1/2}$ decay rate           & $\Gamma (2\pi$ MHz)   & 20.95*          \\ 
$X^2\Sigma^+(\nu=0) \leftarrow A^2\Pi_{1/2}(\nu=0)$ Franck-Condon factor & $\mathcal{F}_{00}$    & 0.96763*    \\ 
\hline\hline
\end{tabular}%
}
\caption{Properties of the experimental platform relevant to this proposal. Molecular properties of $^{24}$MgF obtained from Ref. \cite{Doppelbauer_MgF_2022} except for ones denoted by (*) obtained from Ref. \cite{Norrgard_MgF_decay_branching_2023}. Nearest neighbor distance is set by laser wavelength and configuration.}
\label{tab:properties}
\end{table*}

\section{Experimental Parameter Space and Phase Diagram}\label{sec:properties}
Although the formalism described above is applicable to any $^2\Sigma$ molecule, we present magnesium fluoride (MgF) as the molecule of choice for this proposal. This molecule is ideally suited for quantum gas microscopy due to its light mass, strong (nearly-) cycling laser transition at 359~nm and the availability of both bosonic and fermionic isotopes. For the purpose of this work, the statistics of the particle is not relevant since we consider the molecular locations to be static. Hence, we restrict the formalism to only the bosonic isotope $^{24}$MgF with a simpler hyperfine structure. The properties of MgF are collected in Table~\ref{tab:properties}, and relevant level structure for this work is depicted in Fig. \ref{fig:LevelStructure}.  
The near ultra-violet (UV) transition of MgF at $\lambda =$ 359~nm can be used for direct imaging in an optical trap through a high numerical aperture ($NA$) microscope objective. The resolution for such an imaging system is given by $0.61\frac{\lambda}{NA}$ and is expected to allow short spacing between nearest neighbors in a trap and single-site resolution imaging. Here, we assume a spacing of $a = 376$~nm. The geometry of the system can be tuned from 2D to quasi-1D using spatial light modulators.

We begin with MgF molecules in the first rotational state $(N=1,-)$ of the electronic ground state $(X^2\Sigma^+)$ in an optical trap. In the absence of an external field, the molecules are distributed within the hyperfine manifold of the $(N=1,-)$ state (i.e.,~$F=1, 0, 1, 2)$ based on the degeneracy of the magnetic sublevels. By selectively turning off laser sidebands addressing different hyperfine states during the optical pumping cycle, molecular population can be accumulated predominantly in the desired $J = 3/2, F = 2$ manifold. Any remaining population in the other manifolds can be removed via optical pumping. At this point, a near-resonant microwave drive of $\sigma^+$ polarization at a frequency of $\sim31$~GHz and a static magnetic field of magnitude $\sim40$~mG are simultaneously ramped up. In order to adiabatically prepare the system at the desired point of $\Omega_R \sim 350$~kHz and $\epsilon_z \sim 50$~kHz, a small microwave power is needed. This prepares the molecules in a superposition of $\ket{-}_n$ states which are all degenerate. 

\begin{table}[b]
    \centering
    \bgroup
    \def\arraystretch{1.5}
    \begin{tabular}{c c c c}
    \hline\hline
    Region & Detuning & Zeeman shift & Initial molecular state
     \\
    \hline
    I & Red ($\omega < \omega_0$) & $\epsilon_z > 0$ & $\ket{-}_n$ \\
    II & Blue ($\omega > \omega_0$) & $\epsilon_z < 0$ & $\ket{+}_n$ \\
    III & Red ($\omega < \omega_0$) & $\epsilon_z < 0$ & $\ket{+}_n$ \\
    IV & Blue ($\omega > \omega_0$) & $\epsilon_z > 0$ & $\ket{-}_n$ \\
    \hline \hline
    \end{tabular}
    \egroup
    \caption{Table summarizing the type of detuning, sign of the Zeeman shift $\epsilon_z$ and the initial dipolar molecular state required to access regions I, II, III and IV in Fig.~\ref{fig:SOP}.}
    \label{tab:ExpParams}
\end{table}

For nearest neighbor MgF molecules, which have a dipole moment  $d_0 = 2.88\,\mathrm{D}$, with lattice constant $a = 376$ nm the dipolar interaction strength is $V_{dd} = d_0^2/a^3\sim 50$~kHz. The experimental parameter space is described by three quantities: the detuning $\delta$ and the Rabi frequency $\Omega_R$ of the microwave field, and the Zeeman shift $\epsilon_z$ caused by the dc magnetic field. In particular, $\epsilon_z$ can be tuned to bring the dressed states to near degeneracy, thus enhancing the dipolar interactions between molecules (see Fig.~\ref{fig:Quasienergy}). The dipole interaction matrix $\hat {\mathbb{V}}_{dd}$ in Eq.~\eqref{eq:V_dd}, and hence $H_{\text{eff}}$ in Eq.~\eqref{eq:H_eff}, are determined from the experimental parameters ($\delta, \epsilon_z, \Omega_R$). Additionally, we have freedom in choosing the signs of $\delta$, with $\delta> 0 $ corresponding to blue detuning of the microwave and $\delta < 0$ to red detuning,  and the sign of $\epsilon_z$, with $\epsilon_z > 0$ when orienting the $B$-field in the $+z$-direction and $\epsilon_z < 0$ in the $-z$-direction, independently. This expands the range of accessible values of the various spin-1 couplings and onsite terms in $H_{\text{eff}}$, and provides access to four different regions in the phase diagram. These regions are shown in Fig.~\ref{fig:SOP}(a); Table~\ref{tab:ExpParams} summarizes the type of detuning, sign of $\epsilon_z$ and the initial state preparation of the dipolar molecules required to access those regions. 

\begin{figure}
    \centering
    \includegraphics[width=\linewidth]{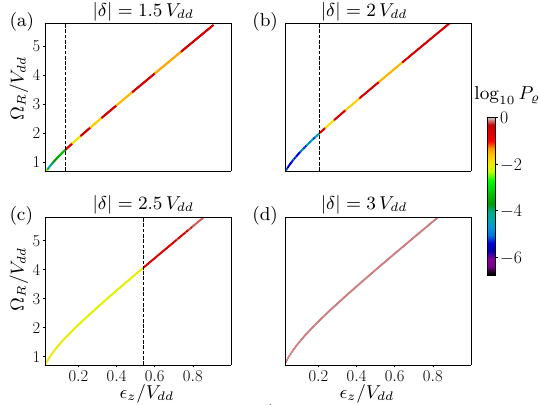}
    \caption{Plots showing $\log_{10}P_\varrho$ for the many-body ground state of $H_{\text{eff}}$ in Eq.~\eqref{eq:H_eff} as functions of $\epsilon_z/V_{dd}$ and $\Omega_R/V_{dd}$ for a red detuned microwave with (a) $|\delta| = 1.5\,V_{dd} \approx 75$ kHz (b) $|\delta| = 2\,V_{dd} \approx 100$ kHz (c)  $|\delta| = 2.5\,V_{dd} \approx 125$ kHz and (d)  $|\delta| = 3\,V_{dd} \approx 150$ kHz. For each coordinate $(\epsilon_z/V_{dd}, \Omega_R/V_{dd})$, the color represents the value of $\log_{10}P_\varrho$. The dashed black line in (a)-(c) represents the Haldane to topologically trivial phase boundary, with the Haldane (topologically trivial) phase to the left (right) of the dashed line. In (d), the entire phase is topologically trivial. The DMRG parameters are the same as in Fig.~\ref{fig:EntSpectrum}.}
    \label{fig:ExpLogPlambda}
\end{figure}

\begin{figure}
    \centering
    \includegraphics[width=\linewidth]{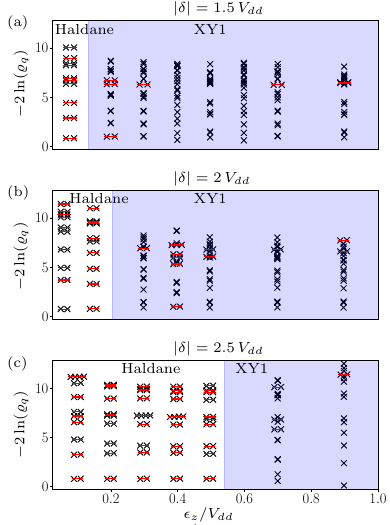}
    \caption{Entanglement spectrum of the many-body ground state of $H_{\text{eff}}$ in Eq.~\eqref{eq:H_eff} as a function of $\epsilon_z/V_{dd}$ along the curves shown in Fig.\ref{fig:ExpLogPlambda}(a)-(c). 
    For degenerate Schmidt values, the crosses are slightly shifted horizontally to indicate multiplicity and groups of degenerate Schmidt values are joined by the red line. The color scheme for the Haldane and XY phases and DMRG parameters are the same as in Fig.~\ref{fig:EntSpectrum}.}
    \label{fig:ExpEntSpectrum}
\end{figure}

Region I is most promising in terms of realizing the Haldane phase. Note that, using the experimental parameters, we simulate the full $H_{\text{eff}}$ in Eq.~\eqref{eq:H_eff} including $\mathbb{V}_{\text{SU(3)}}$. We expect deviations in the phase boundaries due to $\mathbb{V}_{\text{SU(3)}}$ even when the transverse field $h^z_i = 0$. Additionally, $\mathbb{V}_{\text{SU(3)}}$ breaks T-reversal and $\pi$-rotational symmetry, but preserves bond-centered inversion symmetry as shown in Appendix \ref{app:VSU3}. Hence, while this destroys the spin-$\frac12$ edge states as well as the $\mathbb Z_2\times \mathbb Z_2$ classification of the Haldane phase, the phase is still stabilized due to the presence of bond-centered inversion, resulting in a doubly-degenerate entanglement spectrum and vanishing $P_\varrho$.

 In Fig.~\ref{fig:ExpLogPlambda} we show the phase diagram using the pseudo order parameter $P_\varrho$ for the many-body ground state of $H_{\text{eff}}$ in Eq.~\eqref{eq:H_eff}, as a function of experimental parameters $\epsilon_z/V_{dd}$ and $\Omega_R/V_{dd}$, and for a red detuned microwave, where the color of data point represents the value of $\log_{10}P_\varrho$. The dashed black line in Fig.~\ref{fig:ExpLogPlambda}(a)-(c) represents the boundary between the Haldane (left) and topologically trivial phases (right), which we mark by monitoring the changes in $P_\varrho$. The specific values of $(\epsilon_z/V_{dd}, \Omega_R/V_{dd})$ are chosen from a subset of the data points shown in Fig.~\ref{fig:ExpPhaseDiag}(b) corresponding to the lowest row of data points in the bottom right panel,
 but with different values of $|\delta|$, thus showing the evolution of the phase diagram with $|\delta|$. By increasing $|\delta|/V_{dd}$, we move downwards in Region I and the width of the Haldane phase increases until we completely transition into the AFM phase [Fig.~\ref{fig:ExpLogPlambda}(d)]. Consistent with this, Fig.~\ref{fig:ExpEntSpectrum} depicts the entanglement spectrum $\{-2\ln(\varrho_q)\}$ as a function of $\epsilon_z/V_{dd}$ for $|\delta|$ values corresponding to Fig.~\ref{fig:ExpLogPlambda}(a)-(c) respectively. The double degeneracy throughout the entanglement spectrum is visible in the white region, identified as the Haldane phase. 

\begin{figure}
    \centering
    \includegraphics[width=\linewidth]{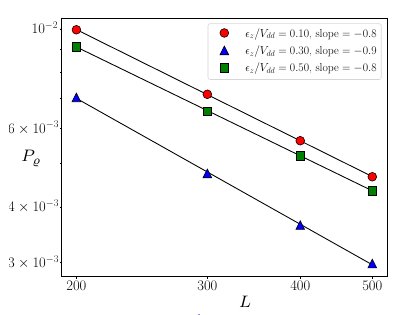}
    \caption{Finite size scaling of $P_\varrho$ for $|\delta| = 2.5\,V_{dd} \approx 125$ kHz with $L=200,300,400,500$ and $\epsilon_z/V_{dd} = 0.1, 0.3, 0.5$ chosen in the Haldane phase, cf. Fig.~\ref{fig:ExpLogPlambda}(c).}
    \label{fig:ExpLogPlambdaScalingL}
\end{figure}

Although the values of $P_\varrho$ are small in this region, they are not exactly zero. This is because of finite-size effects. Focusing on the Haldane region for $|\delta|=2.5V_{dd}$, we performed finite size scaling of $P_\varrho$ for $\epsilon_z/V_{dd} = 0.1, 0.3, 0.5$. We observe a power-law decay with respect to chain size with a decay exponent $\sim 1$ in Fig.~\ref{fig:ExpLogPlambdaScalingL} signifying $P_\varrho \rightarrow 0$ in the thermodynamic limit, and hence confirming that this is indeed the Haldane phase.

Figs.~\ref{fig:ExpPhaseDiag}, \ref{fig:ExpLogPlambda}--\ref{fig:ExpLogPlambdaScalingL} represent the main results of this study.  They show the emergence of the topologically nontrivial Haldane phase within an experimentally accessible parameter space that realizes an effective spin-1 Hamiltonian. The DMRG code using TeNPy library to generate all the data and figures can be found in Ref.~\cite{TeNPy_code}.
 
\section{Conclusions and Outlook}
\label{sec:conclusion}
We have proposed a novel simulation platform for quantum magnetism based on polar molecules. The unique rotational structure and hyperfine couplings found generically among $^2\Sigma$ allows for the creation of microwave dressed states of multiple levels, proportional to the rotational quantum number. By coupling the $N=1,F=2$ states with the $N=0,F=1$ state with circularly polarized microwaves, we can couple three sets of levels simultaneously. Furthermore, we can bring these microwave dressed states to near degeneracy using a static magnetic field. We have shown that the effective dipole-dipole interaction Hamiltonian hosts several quantum phases of matter, notably the SPT Haldane phase. 

We focused on using $^{24}$Mg$^{19}$F molecules trapped in a one-dimensional lattice. Due to the near UV transition wavelength of the molecule, a short lattice spacing of 376~nm can be achieved, thus enhancing the interaction strength $V_{dd}$. The single site resolution microscope images, via the quantum gas microscope, can directly provide spatial correlation functions, such as the string order parameter in $z$-basis. Since the string order parameter carries a signature of the Haldane phase, the single site resolved images can provide evidence of topological symmetry protection in regions where SU(3) terms are negligible.

Our study demonstrates that the Haldane phase is 
robust to SU(3) perturbations and within the reach of reasonable experimental parameters.
When SU(3) terms cannot be neglected, the experimental platform should be able to probe the entanglement entropy or spectrum \cite{PhysRevLett.121.086808}. Particularly, the Renyi entanglement entropy can be accessed via interference measurements on multiple copies \cite{islam2015measuring} or randomized measurement toolbox \cite{elben2023randomized}. Measurement of the entanglement spectrum, on the other hand, is more challenging, especially in analogue quantum simulation platforms, possibly requiring coupling the molecular system to a cavity \cite{PhysRevX.6.041033} or employing quantum learning algorithms \cite{PhysRevLett.127.170501}.

The platform described above is not limited to the realization of the spin-1 XXZ Hamiltonian. By allowing the molecules to quantum tunnel from site to site, one can simulate models of itinerant magnetism. The dimensionality of the simulator can be readily expanded to two-dimensions. Changing the filling fraction can effectively create a model with random couplings, which can host spin glass phases. By coupling higher rotational states, higher effective spin Hamiltonians can be generated. This can be particularly useful for the study of SU(N) magnetism \cite{Mukherjee_2025}. In addition, owing to the larger dipolar interactions of molecules compared to atoms, the dipolar Bose-Hubbard model can be more exhaustively explored in the laboratory.
By changing the molecular isotopologue to $^{25}$Mg$^{19}$F, one can introduce fermions to the platform.

Thus, MgF molecules in optical lattices hold remarkable potential to realize a versatile platform for manifesting a host of correlated and topological many-body states.  The study presented here demonstrates this even for the relatively restricted case of molecules trapped in a one-dimensional lattice, where one may simulate a spin-1 chain, and obtain a remarkably rich set of phases.  This includes the celebrated Haldane phase, in a setting that allows its unusual correlations to be detected.

\begin{acknowledgements}
We acknowledge fruitful discussions with Ana Maria Rey. This work is supported in part by Indiana University Office for Research Development through the Bridge and Emerging Frontiers Programs and the National Science Foundation through Grant Nos. DMR-2533543 and DMR-2531425.
\end{acknowledgements}

\appendix

\section{Microwave dressed states}\label{app:mol_Ham}

Here we derive the dressed states and their energies. We start with the time-dependent Hamiltonian of a single molecule given by Eq.~\eqref{eq:MolHam}. For the electric dipole transition $\ket{N=0,F=1,m_F=n-2}\rightarrow\ket{N=1,F=2,m_F=n-1}$, 
\begin{equation}
    H_n(t) = \begin{pmatrix}
        -\frac{\omega_0}{2} + 2(n-2)\epsilon_z & -\frac{\Omega_R}{x_n}e^{i\omega t} \\
        -\frac{\Omega_R}{x_n}e^{-i\omega t} & \frac{\omega_0}{2} + (n-1)\epsilon_z
    \end{pmatrix}.
\end{equation}
We then perform a unitary transformation to the rotating frame of the molecule, given by $R(t) = \exp\left[i(\mathbb 1 - \sigma_z)\omega t/2\right]$, to obtain the time-independent Hamiltonian in the rotating frame,
\begin{align}
    H_{n,\mathrm{rot}} & =  R(t)H_n(t)R^\dagger(t) - iR(t)\frac{d}{dt}R^\dagger(t) \nonumber \\
    & = \begin{pmatrix}
        -\frac{\omega_0}{2} + 2(n-2)\epsilon_z & -\frac{\Omega_R}{x_n} \\
        -\frac{\Omega_R}{x_n} & \frac{\omega_0}{2} + (n-1)\epsilon_z - \omega
    \end{pmatrix}. \label{eq:H_rot}
\end{align}
Diagonalizing Eq.~\eqref{eq:H_rot}, we obtain Eqs.\eqref{eq:plus} and \eqref{eq:minus} for the dressed states in the lab frame, and Eqs.~\eqref{eq:quasienergy1}-\eqref{eq:quasienergy3} for the dressed state energies, in the main text.

\section{Dipole moment operator and $\hat V^{(0)}_{dd}$}\label{app:Dipole}

The matrix elements of the dipole moment operator in the lab frame $[\vex d(t)]_{nn'}$ are obtained using the electric dipole transition selection rules and the Clebsch-Gordon coefficients for the various transitions. The matrix representation of the vector components $\hat d_\pm(t) = \hat d_x(t) \pm i\hat d_y(t)$ and $\hat d_z(t)$ in the basis of the $\ket{-}_n$ states are,
\begin{align}
     \mathbb d_+(t) & =  -d_0 \begin{pmatrix}
            \frac{a_1b_1}{3}e^{i\omega t} & 0 & 0 \\
            0 & \frac{a_2b_2}{\sqrt{3}}e^{i\omega t} & 0 \\
            -\frac{a_1b_3}{3}e^{-i\omega t} & 0 & \sqrt{\frac{2}{3}}a_3b_3e^{i\omega t}
    \end{pmatrix}  \label{eq:dplusminus}
    \end{align}
and
\begin{align}
     \mathbb d_z(t) & =  d_0\begin{pmatrix}
        0 & \frac{\sqrt{2}}{3}a_1b_2e^{i\omega t} & 0 \\
       \frac{\sqrt{2}}{3}a_1b_2 e^{-i\omega t} & 0 & \frac{a_2b_3}{\sqrt{6}}e^{i\omega t} \\
        0 & \frac{a_2b_3}{\sqrt{6}}e^{-i\omega t} & 0
    \end{pmatrix}. \label{eq:dz}
\end{align}
Plugging Eqs.~\eqref{eq:dplusminus} and \eqref{eq:dz} into Eq.~\eqref{eq:Vdd_op}, and performing a time average we obtain $\hat V_{dd}^{(0)}$ given by the expression,
\begin{align}
    \hat V_{dd}^{(0)} = & \left[-\frac{\hat d_+^{(1)}\otimes \hat d_-^{(-1)} +\hat d_-^{(1)}\otimes \hat d_+^{(-1)}}{4} \right. \nonumber \\
    & \left. \vphantom{\left(\frac{\hat d_+^{(1)}\otimes \hat d_-^{(-1)} +\hat d_-^{(1)}\otimes \hat d_+^{(-1)}}{4}\right)} + \hat d_z^{(1)}\otimes \hat d_z^{(-1)} + \mathrm{h.c.}\right]\frac{(1-3\cos^2\theta_{ij})}{r^3},
\end{align}
where $\hat d_j^{(\pm1)} = \int_0^T \hat d_j(t)e^{\pm i\omega t}\frac{dt}{T}$ are the Fourier components of $\hat d_j(t)$ ($j = \pm, z$). Similar expressions can be obtained for the $\ket{+}_n$ states by simply replacing $b_n\rightarrow a_n$ and $a_n\rightarrow -b_n$.

\section{Symmetry analysis of $\hat V_{\mathrm{SU(3)}}$}\label{app:VSU3}

Here we show the action of $\pi$-rotations $e^{-i\pi S^\alpha}$ on $\hat V_{\mathrm{SU}(3)}$. After performing the minimization procedure described in the main text, Eq.~\eqref{eq:Minimize}, we find that the nonzero one-body interaction terms in $\hat V_{\mathrm{SU}(3)}$ are proportional to the Gell-Mann matrices $\lambda_3$ and $\lambda_8$. Under rotations, $[e^{-i\pi S^{x,y}}, \lambda_3] = \pm i\lambda_5$, $[e^{-i\pi S^{x,y}}, \lambda_8] = \pm i\sqrt{3}\lambda_5$ and $[e^{-i\pi S^z}, \lambda_{3,8}] = 0$. Thus, the one-body terms in $\hat V_{\mathrm{SU}(3)}$ are responsible for breaking $\pi$-rotations about the $x$ and $y$ axes, as well as TR symmetry. However, unlike the transverse field in Eq.~\eqref{eq:H_eff}, the one-body terms in $\hat V_{\mathrm{SU}(3)}$ also break the combination TR$\times e^{-i\pi S^x}$ and TR$\times e^{-i\pi S^y}$. Thus, $\hat V_{\mathrm{SU}(3)}$ spoils the $\mathbb Z_2\times\mathbb Z_2$ protection of the Haldane phase. Nonetheless, the Haldane phase is stabilized by the preserved bond-centered inversion symmetry, as numerically confirmed in the main text. 

\bibliography{refs}

\end{document}